\documentclass[journal,twocolumn,a4paper]{IEEEtran}
\usepackage{epsfig}
\usepackage{epstopdf}
\usepackage{citesort}
\usepackage{graphicx}
\usepackage[small]{caption}
\usepackage{url}
\usepackage{amsmath}
\usepackage{amssymb}
\usepackage{algorithmic}
\usepackage{algorithm}
\usepackage{enumerate}
\usepackage{multirow}
\usepackage{tikz}
\usepackage{setspace}
\usepackage{subfigure}
\usepackage{color}

\newtheorem{theorem}{\bf Theorem}
\newtheorem{proposition}{\bf Proposition}
\newtheorem{lemma}{\bf Lemma}

\newtheorem{definition}{\bf Definition}
\newtheorem{remark}{Remark}
\newtheorem{property}{\textit{Property}}

\usepackage{rotating} % adde by Wayes Tushar
\usepackage{graphicx} % added by Wayes Tushar
\newcommand{\Rmnum}[1]{\expandafter\@slowromancap\romannumeral #1@}
\makeatother
\begin{document}
\title{Prioritizing Consumers in Smart Grid: A Game Theoretic Approach}
\author{Wayes~Tushar,~\IEEEmembership{Member,~IEEE,}
        Jian~A.~Zhang,~\IEEEmembership{Senior~Member,~IEEE,}
         David~B.~Smith,~\IEEEmembership{Member,~IEEE},~H.~Vincent~Poor,~\IEEEmembership{Fellow,~IEEE}      and Sylvie Thi{\'{e}}baux\vspace{-1.1cm}% <-this % stops a space
\thanks{W. Tushar is with the Engineering Systems and Design Pillar at Singapore University of Technology and Design (SUTD), Singapore 138682. (e-mail: wayes\_tushar@sutd.edu.sg).}
\thanks{D. B. Smith and S. Thi{\'{e}}baux  are with the College of Engineering and Computer Science at the Australian National University (ANU), ACT, Australia. They are also with NICTA, Canberra, Australia. (e-mail: \{sylvie.thiebaux,~david.smith\}@nicta.com.au).}
\thanks{J.~A.~Zhang is with CSIRO Computational Informatics, Marsfield, NSW, Australia~(e-mail:~andrew.zhang@csiro.au).}
\thanks{H. V. Poor is with the School of Engineering and Applied Science, Princeton University,
Princeton, NJ, USA~(e-mail:~poor@princeton.edu).}
\thanks{This work was done when W. Tushar was a NICTA affiliated Ph.D. student at ANU, Australia. NICTA is funded by the Australian Government as
represented by the Department of Broadband, Communications and the
Digital Economy and the Australian Research Council through the ICT
Centre of Excellence program.}
}
\IEEEoverridecommandlockouts
\maketitle
\begin{abstract}
This paper proposes an energy management technique for a
consumer-to-grid system in smart grid. The benefit to consumers is
made the primary concern to encourage consumers to participate voluntarily
in energy trading with the central power station (CPS) in situations of energy deficiency. A novel system model motivating energy trading under the goal of social optimality is proposed.
A single-leader multiple-follower Stackelberg game is then studied to model the interactions
between the CPS and a number of energy consumers (ECs), and to find optimal distributed solutions for the optimization problem based on the system model.
The CPS is considered as a leader seeking to minimize its total cost
of buying energy from the ECs, and the ECs are the followers who
decide on how much energy they will sell to the CPS for
maximizing their utilities. It is shown that the game, which can be implemented distributedly, possesses a socially optimal solution, in which
the benefits-sum to all consumers is maximized, as the total cost to the CPS is minimized. Numerical analysis confirms the effectiveness of the game.
\end{abstract}

%\vspace{-12pt}
 \setcounter{page}{1}
\section{Introduction}
%\vspace{-3pt}
\IEEEPARstart{A} key element of smart grid implementation is the enabling
of consumers to participate by encouraging them to provide ancillary
services to the main power grid~\cite{2011IEEE-JCST_Fang}. The
development of new energy management applications and services,
based on consumers' active participation, can help leverage the
technology and capability upgrades available from the smart grid ~\cite{2011IEEE-JCST_Fang}.
%through its advanced infrastructure

In a constrained energy market, the engagement of consumers in
energy management can greatly enhance the grid's reliability, and
significantly improve the social benefit of the overall
system~\cite{2010-JEnergy_Walawalkar}. For instance, a study by
McKinsey \& Company shows that $10-15$ billion US dollars (USD) in
annual benefit can be achieved from large-scale USA-wide active
participation of all customers in energy management
programs~\cite{2010-JEnergy_Walawalkar}. Consequently, energy
management research, in the context of smart grid, has received
considerable attention recently, as can be seen from a large amount of work reviewed in~\cite{2011IEEE-JCST_Fang}.
However, one of the key challenges for successful energy management
in smart grids is to motivate consumers to actively and
voluntarily participate in such management programs. If the %energy
consumers are not interested in actively taking part in energy
management, the benefits of smart grid will not be fully
realized~\cite{2011IEEE-ISGT_Liu}. Therefore, to make the consumers
an integral part of any energy management scheme, the design of the
scheme needs to be consumer-centric~\cite{2011IEEE-ISGT_Liu},
whereby the main recipients of smart grid benefits %should be
are energy consumers as both buyers from, and sellers to, the energy grid.

In this paper, a consumer-centric energy management scheme is
proposed for a consumer-to-grid system that gives significant
benefit to consumers who actively participate in the smart grid. The
idea of consumer-centric smart grid (CCSG) was  first introduced in
~\cite{2011IEEE-ISGT_Liu}. Further, in  \cite{2010IEEE-CEnergyCon_Zafar}, customer
domain analysis of smart grid is studied along with the tasks
arising in this domain. Our energy management scheme in this paper complements the existing work on CCSG by proposing a discriminate pricing strategy to encourage as many energy consumers (ECs) as possible to participate in energy trading with the central unit. In the proposed pricing mechanism, ECs with smaller surplus energy may expect higher unit selling price and the price is adaptive to the number of participating ECs and their offered energy for sale. At the same time, our scheme is also designed to minimize the total purchasing cost for the central power station (CPS). The work presented in this paper significantly extends our previous work in \cite{2013-CICC_Tushar}. It provides an improved and generalized system model, detailed performance analysis of the solution based on the model, and more comprehensive simulation results.

The main contributions of this paper are as follows. 1) A general system model is proposed for facilitating consumer-centric energy management. Novel utility and cost models are proposed to enable discriminate pricing mechanisms. These models achieve a good balance in reflecting practical requirements and providing mathematical tractability; 2) A single-leader multiple-follower Stackelberg game is proposed to solve the above energy management problem by enabling decentralized decision making through limited interaction between the CPS and the ECs; 3) The optimality and the convergence of the proposed algorithm based on the Stackelberg game are proven; and 4) Insights are obtained for the choice of parameters in the system model through both analytical and numerical results.

The rest of the paper is organized as follows. The system model and the optimization problem are presented in Section~\ref{system-model-ch-4}. The proposal for an energy management game to perform this optimization is
described in Section~\ref{game-formulation-ch-4}. The properties
of the game are discussed in Section~\ref{game-property-ch-4}.
Section~\ref{algorithm-ch-4} describes an algorithm to achieve social optimality and Section~\ref{numerical-results-ch-4} gives numerical results. Finally, some concluding remarks are made in Section~\ref{conclusion-ch-4}.
%\vspace{-3pt}
\section{System Model}\label{system-model-ch-4}

Consider a smart grid network that consists of a CPS and multiple ECs. Here, the CPS refers
to a power generating unit that is connected to the ECs of the network by means of power lines, and ECs are the energy entities such as
electric vehicles (EVs), solar and wind farms, smart homes and bio-gas plants, which have energy storage devices (batteries) and communication devices such as smart meters for communicating with the CPS ~\cite{2010IEEE-JTSG_Rad}. Each EC
may represent a group of similar energy customers of the smart grid acting as a single entity. Due to the massive demands of consumers at peak hours, the CPS may be unable to meet the energy demands. Buying energy from ECs can be more cost efficient than setting up expensive generators or bulk capacitors for meeting excess
needs. ECs can voluntarily take part in trading their excess energy with the CPS with appropriate incentives. It is noted that although we mainly target demand management for peak hours in this paper, it is straightforward to extend the proposed scheme to other situations such as during power outages and emergencies, whenever the CPS is unable to meet the demands of consumers.

In this paper, we consider per time slot based energy management, to adapt to the variation of energy usage in a day. For example, the peak hours' operation can be divided into multiple time-slots of 30 minutes each~\cite{2012-AEMC_Pierce}. One main assumption based on the per time slot model is that the CPS is not interested in buying more energy than the goal it sets in advance. This assumption is necessary for making the proposed scheme work efficiently. The proposed scheme can be repeatedly applied over multiple continuous time slots, like an online game, with updated parameters based on the results in the previous time slot and updated participating ECs. The CPS and the ECs can also set their parameters according to statistical prediction models to achieve some benefit similar to that of arbitraging between various time slots. For example, the CPS may seek more than what it actually needs in a time slot, should it see the benefit of doing so. Hence, many parameters to be defined for the system model are time slot dependent, and they can vary from time slot to time slot, along with such dependence for the decision variables output of the proposed energy management scheme.

Let us consider $N$ ECs in a set
$\mathcal{N}$ in the smart grid network, which are participating in energy trading with the CPS. At a particular time slot of
energy deficiency $E_{\text{def}}$, EC $n$ has an amount of energy $E_n$
available to sell to the CPS. $E_n$ may be
different for different $n$ based on parameters such as the type of EC, the
current weather (e.g., a solar farm may wish to sell a large amount
of energy on a sunny day compared to other cloudy or rainy days) and
the capacity of the storage device. The amount of energy
$E_{\text{def}}$ required by the CPS is assumed to
be fixed, and hence the energy supplied by all ECs to the CPS needs to satisfy the constraint
\begin{equation}
\label{eqn-supply-constraint-ch-4} \sum_ne_n\leq E_{\text{def}}; e_n\leq
E_n,~\forall n\in\mathcal{N},
\end{equation}
where $e_n$ is the energy supplied by EC $n$. The use of $\sum_ne_n\leq E_{\text{def}}$ instead of $\sum_ne_n= E_{\text{def}}$ is based on the fact that this is a ``best-effort'' activity and it is not always guaranteed to achieve $E_{\text{def}}$.

Now, we want to design an energy management scheme to achieve \textit{social optimality} in the energy trading. Social optimality means that all players can benefit from the energy trading to maximize the social welfare, which represents the sum of all ECs' and CPS's utilities, rather than an individual`s benefit. Achieving social optimality implies that 1) every EC with energy surplus can participate in energy trading and is motivated to do so; 2) each EC can optimize its benefit when social welfare is maximized. Such benefit may be smaller than the maximum that each EC can individually achieve without considering the social welfare; and 3) the overall energy purchasing cost can be controlled and minimized to benefit all consumers. Hence the scheme should allow and encourage as many ECs as possible to participate in energy trading by balancing their expectations and returns, rather than overly emphasizing individual's benefit. Such optimality will ultimately reward all ECs as both energy consumers and providers. As to be seen later, the social optimality here matches well with the social optimality in the \textit{Generalized Nash Game}. Next, we present three models, which are designed to encourage more ECs to participate in energy trading and minimize energy shortage and purchasing cost, and ultimately to benefit all consumers, and achieve such social optimality.

\subsection{Unit Price Model}

More trading ECs can lead to better completion of the purchasing target and more savings on buying cost. However, not all the ECs are interested in trading energy with the CPS if the benefit is not attractive. This could particularly happen to numerous ECs with smaller $E_n$ whose expected return can be small under a feed-in tariff (FIT) scheme. In this case, ECs would store the energy, due to uncertainty, rather than selling it. To encourage as many ECs as possible to participate, we want the CPS to provide different incentives to different ECs, depending mainly on their energy available for sale and also on their preferences. This is achieved through the unit energy price (price per unit of energy), $p_n$, that the CPS pays to EC $n$ for its offered energy $e_n$. In our scheme, $p_n$ can be different for different ECs, and these are adaptively determined by the CPS during the trading process with the ECs, through their supplied energy as to be seen later. Note that the current grid system does not allow discriminate pricing among consumers.
However, real-time pricing is an envisaged addition to future smart
grids ~\cite{2010-WEIS_Anderson} and an example of this is found in standard FIT schemes~\cite{2010-NREL_Couture}.

The CPS wants to minimize its total cost of purchasing energy so that it can sell the energy to its consumers at a cheaper rate, which in turn will benefit all the consumers. Therefore, we introduce
a ``total unit energy price" parameter $P=\sum_n p_n$, analogous to the ``total cost per unit production" widely used in economics~\cite{2010-Book_Farris}. Here, the parameter $P$ is used by the CPS to control the total purchasing cost. As to be seen in Section \ref{sec-cpsdec}, $P$ scales a set of normalized prices to generate the unit energy prices $p_n$, and hence the total direct energy purchasing cost (the sum of the product $e_np_n$) is linearly proportional to $P$. Such a $P$ will also be used to determine the initial $p_n$ as $P/N$ in our proposed scheme. The parameter $P$ is fixed for each time slot, and can be determined by
the CPS using any real-time price estimator such as that proposed
in ~\cite{2008IEEE-JTPS_Yun}.

At the same time, we also require $p_{\text{min}}\leq p_n\leq p_{\text{max}}$ where $p_{\text{min}}$ and $p_{\text{max}}$ are the minimum and maximum price per unit energy. The lower bound $p_{\text{min}}$ is used to prevent an EC from being deterred from energy trading. The upper bound $p_{\text{max}}$ can be used to prevent the CPS from allowing a too large of a $p_n$, and hence this reduces the overall purchasing cost. Their values are in the range of $0\leq p_{\text{min}},p_{\text{max}}\leq P$, and any interim $p_n$ will be rounded to either $p_{\text{min}}$ or $p_{\text{max}}$ when it is out of this range.

The final price model we have is that the CPS pays $p_n$ to EC $n$
based on its offered energy $e_n$, while maintaining the constraint
\begin{eqnarray}
\sum_n p_n=P,~p_{\text{min}}\leq p_n\leq p_{\text{max}}.
 \label{eqn-price-constraint-ch-4}
\end{eqnarray}

\subsection{Utility Model and ECs' Objectives}\label{problem-formulation-ch-4}

%%\subsection{Objective of the ECs}
%% ---------------------------------------------------------------------------
%\begin{figure}[t!]
%\centering
%\includegraphics[width=0.8\columnwidth]{utility_function.eps}
%\caption{Effect of the amount of supply energy on the utility of a consumer.}
%\label{fig:utility-function}
%\vspace{-5pt}
%\end{figure}
%% ---------------------------------------------------------------------------

In general, each EC's objective is to maximize its own benefit. However, such an objective can only be achieved when social optimality is achieved. Without considering the social optimality, the CPS will very likely disappoint most of ECs when maximizing the benefits of only a limited number of ECs. This can result in a significant reduction in participating ECs, and degrade the performance of energy management. One way of maximizing ECs' benefits under the social optimality constraint is through maximizing a function representing the sum of all ECs' benefits, with an individual EC being able to set its preference in the function. For this purpose, we consider the function as a sum of each individual's utility function.

The $n^\text{th}$ EC's benefit depends on the unit price $p_n$, the supplied energy $e_n$, and the available energy for sale $E_n$. Hence the individual's utility function can be written as $U(e_n, E_n,p_n)$. A good utility function should have the following two properties.
\begin{property}
The utility function is an increasing function of $p_n$ and $e_n$, i.e., $\partial U(e_n,E_n,p_n)\,/\,\partial p_n>0$ and $\partial U(e_n,E_n,p_n)\,/\,\partial e_n>0$.
\end{property}

\begin{property}
The utility function is a concave function of $e_n$, i.e., $\partial^2 U(e_n,E_n,p_n)\,/\,\partial^2 e_n<0$,
which means that the utility can become saturated or even decrease with an excessive $e_n$. This reflects the fact that since a consumer is equipped with a battery with limited capacity, extensive supply of electricity once exceeding a certain limit would risk the depletion of the battery because of the calendar ageing effect \cite{Eddahech20122438} and consequently, decrease the consumer's utility.
\end{property}

%\begin{property}
%The utility function is an increasing function of $E_n$, i.e., $\partial U(e_n,E_n,p_n)\,/\,\partial E_n>0$.
%\textbf{But why?}
%\end{property}

Among many potential utility functions possessing the above properties, we propose to use the following one:
\begin{eqnarray}
U(e_n,E_n,p_n) =p_ne_n+(E_n-c_ne_n)e_n.\label{utility-function-ch-4}
\end{eqnarray}
Here, $p_ne_n$ represents the direct income an EC can receive, and $(En-c_ne_n)e_n$ represents the possible loss where $c_n\geq0$ is a constant that can be chosen to suit different ECs' preferences. Different values of $c_n$ reflect the different negative impacts of extensive supply on an EC's utility. An EC can set a larger $c_n$ if it prefers to sell less. Introducing $E_n$ into the model is to emulate the fact that ECs with different amounts of energy available for sale can tolerate different thresholds of extensive supply, and utility decreases only when $e_n$ exceeds the threshold. Introducing $E_n$ in the form of $E_ne_n$ also allows EC $n$ to decide its offered energy $e_n$ proportional to its available energy $E_n$ in its decision making process as to be seen in Section \ref{sec-decec}. This function possesses the so-called feature of \textit{linearly decreasing marginal benefit} which has been widely adopted in various utility functions~\cite{2010IEEE-CSmartGridComm_Samadi}. With the goal of maximizing the sum of individual's utilities, the common objective of ECs can be represented as
\begin{align}
&\max_{\mathbf{e}}\widetilde{U}(\mathbf{e,E,p})=\max_{e_n, n=1, \cdots, N}\sum_n \big(p_ne_n+(E_n-c_ne_n)e_n\big),\notag \\
&\text{subject to} \sum_ne_n\leq E_{\text{def}},\label{obj-AES-ch-4}
\end{align}
where $\mathbf{e}=[e_1, e_2, \hdots, e_N]^T$, $\mathbf{E}=[E_1, E_2, \hdots, E_N]^T$ and $\mathbf{p}=[p_1, p_2, \hdots, p_N]^T$. That is, EC $n$ chooses $e_n\leq E_n$, to supply to the CPS so as to maximize the sum of utilities in \eqref{obj-AES-ch-4}.
%However, in actual implementations as to be seen later, each EC needs only to consider how to maximize its own utility, under the common coupled constraint of (\ref{eqn-supply-constraint-ch-4}), through implicit coordination by the CPS.

\subsection{Cost Model and the CPS's Objective}

While the objective of an EC is to maximize its utility through
its choice of $e_n$, the CPS wants to minimize its total cost. Although the direct purchasing cost is $\sum_n e_np_n$, we propose to use the following function to better capture the total incurred cost:
\begin{align}
\tilde{L}(\mathbf{p, e}, E_{\text{def}})=&\sum_n(e_np_n^r+a_np_n+\beta_ne_n+b_n)+\notag\\
&\alpha(E_{\text{def}}-\sum_ne_n),
\label{eq-Ltot}
\end{align}
where $e_np_n^r, r>1$ corresponds to the direct cost $e_np_n$ but is weighted by $p_n^{r-1}$, in order to generate discriminate prices for ECs with different $e_n$s; the term $(a_np_n+\beta_ne_n+b_n)$, with $a_n, \beta_n, b_n\geq 0,$ ~$\forall~n\in\mathcal{N}$, accounts for the costs associated with transmission and store of the purchased energy; and $\alpha(E_{\text{def}}-\sum_ne_n),\alpha\geq0$ denotes the cost associated with insufficient energy purchasing, for example, shed load.

For simplicity, we assume $\beta_n=\alpha$ and discard the term $\alpha E_{\text{def}}$ in (\ref{eq-Ltot}) and obtain
\begin{align}
\tilde{L}(\mathbf{p, e})=&\sum_n(e_np_n^r+a_np_n+b_n)=\sum_n L(p_n,e_n),
\label{eqn-lsimp}
\end{align}
where $L(p_n,e_n)\triangleq e_np_n^r+a_np_n+b_n$ denotes the individual cost function for EC $n$. Note that it will become clear in Section \ref{sec-cpsdec} that such simplification has little influence on our scheme. The analysis for the scheme will be presented from Section \ref{game-formulation-ch-4} to \ref{algorithm-ch-4}.

Now the objective of the CPS can be formally presented as
\begin{eqnarray}
&&\min_{\mathbf{p}}\tilde{L}(\mathbf{p,e})=\min_{{p_n, n=1,\cdots, N}}\sum_n(e_np_n^r+a_np_n+b_n),\notag
\\&&\text{subject to}~\sum_n p_n=
P,~p_{\text{min}}\leq p_n\leq p_{\text{max}},~\forall n.\label{obj-CEU-ch-4}
\end{eqnarray}

\subsection{Optimization Problem}

The optimization problems in \eqref{obj-AES-ch-4} and
\eqref{obj-CEU-ch-4} are connected by $p_n$ and $e_n$. The CPS can find solutions for both problems by jointly optimizing \eqref{obj-AES-ch-4}
and \eqref{obj-CEU-ch-4} in the case when the CPS has full control
over the decision making processes of the ECs. However, in practice,
the CPS does not have any direct control over the ECs' decisions as
these are made by each customer~\cite{2012IEEE-JTSG_Wu}, and parameters such as $E_n$ and $c_n$ can be unknown to the CPS.
Therefore, a decentralized control mechanism is required for the ECs
to decide on the energy they sell to the CPS to realize the optimization in
\eqref{obj-AES-ch-4}. The mechanism also needs to successfully
capture the interaction between the ECs and the decision making of
the CPS for prescribed energy trading. We propose such an energy
management mechanism, using game theory, in the next section.
%\vspace{-4pt}

\section{Non-Cooperative Game Formulation}\label{game-formulation-ch-4}
To decide on energy trading parameters, a single-leader multiple-follower Stackelberg game
~\cite{2012IEEE_JTSG_Tushar} is proposed to study the interaction between the CPS and the ECs. In the proposed
Stackelberg game, the CPS is the leader of the game, which decides on unit energy price $p_n$, within constraint
\eqref{eqn-price-constraint-ch-4}, to be paid to the EC $n$ for its offered energy $e_n$. Each EC $n\in\mathcal{N}$ is a
follower that plays a generalized Nash
game~\cite{2007-J4OR_Facchinei} with other ECs in the network to
decide on the amount of energy it will sell to the CPS, within
constraint \eqref{eqn-supply-constraint-ch-4} in response to the
price $p_n$. Note that this is not just a Nash game, where each player needs to maximize it's own utility, but a generalized Nash game, where all players need to maximize the sum of their utilities and hence the social welfare, due to the presence of the common coupled constraint in \eqref{eqn-supply-constraint-ch-4}. Thus, the Stackelberg game can be formally defined by its
strategic form as
\begin{eqnarray}
\Gamma =
\{(\mathcal{N}\cup\{\text{CPS}\}),\{\mathbf{E_n}\}_{n\in\mathcal{N}},
\widetilde{U},\tilde{L},\mathbf{p}\},\label{formal-game-ch-4}
\end{eqnarray}
where
\begin{itemize}
\item $\left(\mathcal{N}\cup\{\text{CPS}\}\right)$ is the total set
of players in the game, where $\mathcal{N}$ is the set of followers who act in response to the action taken by the leader of the game in set $\mathcal\{{\text{CPS}}\}$;
\item $\mathbf{E_n}$ is the strategy vector of
each EC $n\in\mathcal{N}$ satisfying the constraint in
\eqref{eqn-supply-constraint-ch-4}, i.e., $\sum_n e_n\leq
E_{\text{def}},~e_n\in\mathbf{E_n},~\forall n\in\mathcal{N}$;
\item $\widetilde{U}$
is the objective function that each EC $n$ wants to maximize.
$\tilde{L}$ is the objective function of the CPS; and
\item $\mathbf{p}$
is the strategy vector of the CPS.
\end{itemize}

It is assumed that the ECs maintain their privacy,
and do not inform each other of the amount of energy they offer to the CPS. This leads to a non-cooperative
Stackelberg game in which the followers do not communicate with each other, but
they may interact with the leader by controled
signaling through smart meters
\cite{2010IEEE-JTSG_Rad}. For example, the CPS can send a single bit to EC $n$ if its offered energy is beyond the constraint in
\eqref{eqn-supply-constraint-ch-4} given the energy offered by other
ECs in the network. Importantly, in this game, the decision
making process of an EC $n$ depends not only on its own strategy
but also on the strategy of other ECs in the network via
\eqref{eqn-supply-constraint-ch-4}. Thus, the generalized Nash game
amongst the ECs, to decide on the amount of energy to be supplied to
the CPS by each EC $n$, is a jointly convex generalized Nash
equilibrium problem (GNEP), in which the ECs' actions are coupled
solely by constraint
\eqref{eqn-supply-constraint-ch-4}~\cite{2007-J4OR_Facchinei}. The
solution of a GNEP is the generalized Nash equilibrium
(GNE)~\cite{2007-J4OR_Facchinei}.

The game is initiated as soon as the ECs in the network start
playing a GNEP for a price $p_n=p,~\forall n\in\mathcal{N}$,
announced by the CPS. The ECs play the GNEP and offer, according to their GNE, the
amount of energy they wish to sell to the CPS at price
$p$. For a similar price $p$, each EC receives a similar
incentive, and thus the offered energies reflect the ECs' supply capacities. With such insight into the capacity of each EC's energy
supply, the CPS decides on its optimal price vector
$\mathbf{p^*}=[p_1^*, p_2^*, ..., p_N^*]^T$ to pay the ECs by
solving the constrained optimization problem in \eqref{obj-CEU-ch-4}
using convex optimization~\cite{2010IEEE-JTSG_Rad}.
Thereafter, as soon as the ECs decide on their GNE energy vector
$\mathbf{e^*} = [e_1^*, e_2^*, ..., e_N^*]^T$, after playing the
GNEP for the optimal price vector $\mathbf{p^*}$, the proposed Stackelberg game
reaches equilibrium. From here on, the solution of the
proposed Stackelberg game $\left(\mathbf{e^*}, \mathbf{p^*}\right)$ will be
referred to as an energy management equilibrium solution (EMES) in
which the CPS will decide on an optimized price vector
$\mathbf{p^*}$ to pay to the ECs in the network, and the ECs will
agree on a GNE energy vector $\mathbf{e^*}$ to be supplied to the
CPS for the given $\mathbf{p^*}$.

\begin{definition}\label{definition-1-ch-4} Consider the Stackelberg game $\Gamma =
\{(\mathcal{N}\cup\{\text{CPS}\}),\{\mathbf{E_n}\}_{n\in\mathcal{N}},
\widetilde{U},\tilde{L},\mathbf{p}\}$ where $\widetilde{U}$ and
$\tilde{L}$ are defined by \eqref{obj-AES-ch-4} and
\eqref{obj-CEU-ch-4} respectively. A set of strategies
$(\mathbf{e^*},\mathbf{p^*})$ constitute the EMES of this game if
and only if it satisfies the following set of inequalities:
\begin{eqnarray}
\widetilde{U}(e_n^*,\mathbf{e_{-n}^*},\mathbf{E},\mathbf{p})\geq
\widetilde{U}(e_n,\mathbf{e_{-n}^*},\mathbf{E},\mathbf{p}),\nonumber\\
\forall e_n\in\mathbf{e},~n\in\mathcal{N},\sum_ne_n\leq
E_{\text{def}},\label{cond-equilirbium_1-ch-4}
\end{eqnarray}
and
\begin{eqnarray}
L(p_n^*,\mathbf{p_{-n}^*})\leq
L(p_n,\mathbf{p_{-n}^*}),\nonumber\\~\forall n\in\mathcal{N},\forall
p_n\in\mathbf{p}, p_{\text{min}}\leq p_n\leq
p_{\text{max}},\label{cond-equilirbium_2-ch-4}
\end{eqnarray}
where $\mathbf{e_{-n}}$ is the GNE energy vector of all the ECs in
the set $\mathcal{N}\setminus\lbrace n\rbrace$ which denotes the new set after removing EC $n$ from $\mathcal{N}$, $\mathbf{p_{-n}}$
is the price vector set by the CPS for all the ECs in the set
$\mathcal{N}\setminus\lbrace n\rbrace$, and $\mathbf{E}$ is the set of strategies of all ECs satisfying~\eqref{eqn-supply-constraint-ch-4}.
\end{definition}
Thus, at EMES, no EC can improve its utility by deviating from its
EMES strategy provided all other ECs are playing their EMES
strategies. Similarly, deviation from EMES price $p_n^*,~\forall
n\in\mathcal{N}$, cannot lower the total cost for the CPS once the
Stackelberg game reaches the EMES.

\section{Properties of the Game}\label{game-property-ch-4}

\subsection{Existence of Equilibrium}
In a non-cooperative game, the existence of an equilibrium (in pure
strategies) is not always
guaranteed~\cite{1999Book_Dynamicgame-Basar}. Moreover, for
consumer-centric smart grids, it is important that the solution be
beneficial for all the consumers in the
network~\cite{2011IEEE-ISGT_Liu}. Therefore, the existence and
optimality of a solution of the proposed Stackelberg game needs to be
determined.
%\subsection{Existence and optimality of the solution}

\begin{lemma}\label{lemma-1-ch-4} A solution exists for the proposed Stackelberg game if the GNEP amongst the ECs in the smart grid network constitutes a generalized Nash equilibrium. The solution will be socially optimal if the GNE of the GNEP is also socially optimal.
\end{lemma}

\begin{IEEEproof}
As the game is formulated, the proposed Stackelberg game reaches the EMES as
soon as the ECs in the network agree on a GNE energy vector to be
supplied to the CPS in response to the optimized price vector
$\mathbf{p^*}$ set by the CPS. The cost function for the CPS in
\eqref{obj-CEU-ch-4} is a strictly convex function, and thus, a
unique solution always exists for the CPS's optimization problem in
choosing $p_n$ for the EC $n,~\forall
n\in\mathcal{N}$,~\cite{2005-BOOK_Dattorro} because the optimization is done on a convex set. Therefore, the
existence of a solution for the GNEP among the ECs, for this unique
price vector, would guarantee the existence of an EMES in the
proposed Stackelberg game. Similarly, the solution will be a socially optimal solution
if the GNE of the GNEP amongst the ECs leads to a socially optimal GNE.
\end{IEEEproof}

To investigate the existence and the optimality of the solution of
the proposed GNEP, first, we formulate the GNEP as a variational
inequality (VI) problem
$\text{VI}(\mathbf{E,F})$~\cite{2011-CWWWC_Arganda}, which is
essentially to determine a vector
$\mathbf{e^*}\in\mathbf{E}\subset\mathbb{R}^n$, such that
\begin{align}
\langle\mathbf{F(e^*),e-e^*}\rangle\geq0, \mathbf{e}\in\mathbf{E}, \notag
\end{align}
where $\mathbf{e}= [e_1,e_2,\cdots,e_N]^T$,
$\mathbf{F}=-\nabla_e
U(e_n,E_n,p_n)$, and $\langle\mathbf{x,y}\rangle$ denotes the inner product of $\mathbf{x}$ and $\mathbf{y}$.

The solution of the $\text{VI}(\mathbf{E,F})$ is a variational
equilibrium (VE)~\cite{2007-J4OR_Facchinei}. In the proposed
scheme, we are particularly interested in showing the
existence and efficiency of the VE. This is because the
proposed GNEP is a jointly convex GNEP due to the coupled constraint
\eqref{eqn-supply-constraint-ch-4}, and hence the VE is the socially
optimal solution among all the GNEs~\cite{2007-J4OR_Facchinei}.
Therefore, in designing a socially optimal consumer-centric energy
management scheme, it is our primary interest to demonstrate the
existence and efficiency of a VE solution. In the rest of this
paper, we will use the terms ``GNEP" and ``variational inequality"
interchangeably.

\begin{theorem}\label{theorem-2-ch-4} The consumers' game amongst the ECs in response
to the CPS's decision vector, i.e., the price vector, possesses a
socially optimal variational equilibrium.
\end{theorem}
\begin{IEEEproof}
It can be proven by showing the existence and uniqueness of the VE through the pseudo-gradient of the utility function in (\ref{obj-AES-ch-4}). The interested reader is referred to \cite{2013-CICC_Tushar}, where the complete proof is provided.
\end{IEEEproof}
\begin{remark}\label{remarl-1-ch-4} From Theorem~\ref{theorem-2-ch-4}, the GNEP
among the ECs in response to the unique optimized price paid to them
by the CPS admits a socially optimal solution. As a consequence, as
proved in Lemma~\ref{lemma-1-ch-4}, the proposed Stackelberg game of consumer-centric energy management
possesses a socially optimal solution.
\end{remark}
%\vspace{-5pt}
\subsection{Decision Making Process}
For a clear understanding of the decision making process of the
players at EMES, we formulate Karush-Kuhn-Tucker (KKT) conditions,
using the method of Lagrange
multipliers~\cite{1995BOOK-NonProg_Bertsekas}, for both ECs' and CPS's
optimization problems.

\subsubsection{ ECs' decisions}\label{sec-decec}
%The KKT condition for ECs' GNEP, i.e., variational inequality
%problem, is given by~\cite{2007-J4OR_Facchinei}
%\begin{eqnarray}
%\mathbf{F(e)}+\xi\nabla_e(\sum_ne_n-E_{\text{def}})=0,\nonumber\\
%\xi(\sum_ne_n-E_{\text{def}})=0,~\xi\geq 0,\label{KKT_AES-ch-4}
%\end{eqnarray}
%where $\mathbf{F}$ is defined in \eqref{eqn-proof-21-ch-4} and $\xi$
%is the Lagrange multiplier. It is important to note that the same
%multiplier $\xi$ is used for all the ECs in the network in
%\eqref{KKT_AES-ch-4} (i.e., $\xi_n = \xi,~\forall n\in\mathcal{N}$).
%This is due to the fact that the GNEP amongst the ECs is a jointly
%convex GNEP, and the solution of this jointly convex GNEP is a VE
%\cite{2007-J4OR_Facchinei}.
The solution of the KKT condition ~\cite{2007-J4OR_Facchinei} for any EC $n$'s GNEP is
\begin{eqnarray}
E_n - 2c_ne_n + p_n-\xi = 0,~\xi\geq0.\label{KKT-AES-2-ch-4}
\end{eqnarray}
This leads to
\begin{eqnarray}
e_n=(E_n + p_n-\xi)/(2c_n)\leq(E_n+p_n)/(2c_n).\label{KKT-AES-3-ch-4}
\end{eqnarray}

Equation (\ref{KKT-AES-3-ch-4}) indicates that at equilibrium, the energy EC $n$ offering for sale is proportional to the unit energy price $p_n$ and its available energy for sale $E_n$, scaled by the constant $c_n$. If an EC prefers to sell more energy, it can choose a smaller $c_n$.
\subsubsection{CPS's decision}\label{sec-cpsdec}
The Lagrangian for CPS's optimization in (\ref{obj-CEU-ch-4}) is given by
\begin{eqnarray}
\Omega =
\sum_n\left(e_np_n^r+a_np_n+b_n\right)+\lambda(P-\sum_np_n),\label{eqn-1-ch-4}
\end{eqnarray}
where $\lambda$ is the Lagrange multiplier. From \eqref{eqn-1-ch-4}, we get
\begin{align}
&\partial \Omega\,/\,\partial
p_n=re_np_n^{r-1}+a_n-\lambda=0,\label{eqn-2-ch-4}\\
&\partial \Omega\,/\,\partial \lambda = P-\sum_np_n=0.\notag
\end{align}
Now assuming the associated costs are the same for all the ECs, i.e.,
$a_n=a$ and $b_n = b$ for any $n$, from
\eqref{eqn-2-ch-4} we get
\begin{eqnarray}
\left(\frac{p_{n_1}}{p_{n_2}}\right)^{r-1} = \frac{e_{{n_2}}}{e_{n_1}},~ n_1, n_2\in[1,N],~ n_1\neq n_2.
\label{eqn-unitprice}
\end{eqnarray}
Thus, if $N$ ECs are connected to the CPS and play a GNEP to decide
on their amounts of energy to be sold to the CPS, at equilibrium the
unit energy price paid to an EC by the CPS is inversely proportional to the energy it offers. Within the constraint of $p_{\text{min}}\leq p_n\leq p_{\text{max}}$ for all $n$, the unit price $p_n$ can be computed as
\begin{align}
p_n=\frac{e_n^{\frac{1}{1-r}}}{\sum_{n=1}^N e_n^{\frac{1}{1-r}}}\,P.
\label{eqn-pnP}
\end{align}
When an obtained $p_n$ is out of the range $[p_{\text{min}},p_{\text{max}}]$, the $p_n$ will be rounded to either $p_{\text{min}}$ or $p_{\text{max}}$.

Equation (\ref{eqn-pnP}) shows that the unit energy price is linearly proportional to $P$, and thus is the direct purchasing cost $\sum_np_ne_n$. It also indicates that the discriminate pricing mechanism can be flexibly realized by setting different values of $r$ for different motivating strategies. The smaller the $r$, the larger the difference between the unit energy prices $p_n$. Hence, for social optimality, in which all the ECs participate in energy trading
with the CPS to their benefit, the consumers with less energy to sell are given greater incentives to play the
game.

According to the analysis above, we can see that a similar rule to (\ref{eqn-pnP}) for determining $p_n$ can be obtained when we replace the cost function (\ref{eqn-lsimp}) with (\ref{eq-Ltot}) in the CPS's decision process. This shows that the major theoretical results derived in this paper can be directly translated to those for the more general cost function in (\ref{eq-Ltot}). Hence the simplification from (\ref{eq-Ltot}) to (\ref{eqn-lsimp}) does not affect the efficiency of the proposed scheme.
%\vspace{-5pt}
%
\section{Algorithm}\label{algorithm-ch-4}
To reach the EMES of the proposed game, an algorithm is proposed in this section that can be implemented by the CPS and the ECs in a distributed fashion with \emph{limited communication} between one another. We note that the decision making process of the ECs can be modeled as a strongly monotone VI problem as can be seen from Theorem \ref{theorem-2-ch-4}. For this problem, the slack
variable, $\xi_n = E_n-e_n+p_n$, possesses the same value for all
the ECs, i.e., $\xi_n=\xi,~n=1,\cdots,N$, when their choice of supply
amount of energy reaches the VE~\cite{2007-J4OR_Facchinei}. This property is being used by the CPS in the
algorithm to check the convergence of the proposed
GNEP to the VE and inform the ECs about it. Here, a hyperplane projection method, particularly the S-S hyperplane
projection method (SSHPM)~\cite{1999-JSIAMJCP_Solodov}, is used to solve the monotone variational inequality. The CPS decides on its unit energy price to pay to each EC by using any standard convex
optimization technique.

%To reach the EMES of the proposed game, an algorithm is proposed in
%this section that can be implemented by the CPS and the ECs in a
%distributed fashion with limited communication between one another.
%The GNEP of the decision making
%process of the ECs is a strongly monotone variational inequality problem \cite{2013-CICC_Tushar}. Therefore, the solution of the
%energy supply game among the ECs, in response to the price set by the
%CPS, can be obtained by solving such a problem. The CPS decides on its price
%per unit of energy to pay to each EC by using any standard convex
%optimization technique.
% ------------------------------------------------------------------------
\begin{algorithm}[t!]
\caption{Algorithm to reach EMES}
\begin{algorithmic}
\STATE \textbf{\emph{Step-1}}%
\STATE\hspace{1mm}(i)- CPS announces $E_{\text{def}}$ and $P$.%
\STATE\hspace{1mm}(ii)- Each EC $n$ calculates $p_n=P/N$,
and ECs play a GNEP to determine VE energy $e_n$, for $p_n$, using SSHPM.%
\STATE\hspace{1mm}(iii)- Each EC $n$ submits VE
energy to CPS.%
\STATE\hspace{1mm}~(iv)- CPS optimizes \eqref{obj-CEU-ch-4}
using a
standard convex optimization technique~\cite{2004-Book_BOYD}, and determines $p_n=p_n^*~\forall n\in\mathcal{N}$.%
\STATE\hspace{1mm} \textbf{The optimized price vector $\mathbf{p^*}$
is obtained}.%
\STATE\textbf{\emph{Step-2}}%
\STATE\hspace{1mm}(v)- Each EC $n$ receives
$p_n^*$ offered by CPS.%
\STATE\hspace{1mm}(vi)- ECs play a GNEP using SSHPM to
determine the VE energies $e_n^*$ to supply to CPS.
\STATE\hspace{1mm} \textbf{The VE energy vector $\mathbf{e^*}$ for
$\mathbf{p^*}$ is obtained}.%
\STATE\emph{\textbf{The game reaches the EMES.}}%
%\vspace{3mm}%
\STATE \textbf{S-S Hyperlane Projection Method (SSHPM)}%
\STATE 1)~At iteration $k$, each EC $n$ computes the hyperplane
projection $r(e_n^{(k)})$ and updates $e_n^{(k+1)}=r(e_n^{(k)})$.%
\STATE\hspace{2mm} \textbf{if} $r(e_n^{(k)})=0$%
\STATE\hspace{3.5mm} a) EC $n$ determines $e_n=e_n^{(k+1)}$ to
offer to CPS.%
\STATE\hspace{3.5mm} b) EC $n$ sends $\xi_n = E_n-2c_ne_n+p_n$ to
CPS.%
\STATE\hspace{2mm} \textbf{else}%
\STATE\hspace{2mm} EC $n$%
\STATE\hspace{3.5mm} a) Determines the hyperplane $z_n^{(k)}$
and the half space $H_n^{(k)}$ from the projection.%
\STATE\hspace{3.5mm} b) Updates $e_n^{{(k+1)}}$ from the
projection of $e_n^{(k)}$ on $E\cap H_n^{(k)}$.%
\STATE\hspace{3.5mm} c) Determines $e_n=e_n^{(k+1)}$ to offer to CPS.%
\STATE\hspace{3.5mm} d) Sends $\xi_n = E_n-e_n+p_n$ to CPS.%
\STATE\hspace{2mm} \textbf{end if}%
\STATE 2) CPS checks $\xi_n~\forall
n\in\mathcal{N}$.%
\STATE\hspace{2mm} \textbf{if} $\xi_n=\xi~\forall n\in\mathcal{N}$%
\STATE\hspace{3.5mm} CPS informs ECs to end the iterations.%
\STATE\hspace{2mm} \textbf{else}%
\STATE\hspace{3.5mm} CPS acknowledges ECs to update their
offered energy at next iteration $k=k+1$.%
\STATE\hspace{2mm} \textbf{end if}
\STATE \textbf{End of SSHPM}%
\end{algorithmic}
\label{algorithm-1-ch-4}
\vspace{-2pt}
\end{algorithm}

As presented in Algorithm~\ref{algorithm-1-ch-4}, the algorithm is executed in two steps assuming that all the
information exchanges between the CPS and the ECs are done through
two way communication via their smart meters
~\cite{2010IEEE-JTSG_Rad}. It starts with the announcement of the
required energy $E_{\text{def}}$ and the
total unit energy price $P$ by the CPS. In the first step, each EC $n$ in the network assumes its
own equally distributed unit energy price $p_n=P/N$,
and plays a GNEP to decide on the amount of energy it would offer to
the CPS for this price, within constraint
\eqref{eqn-supply-constraint-ch-4}. Knowing the offered energy from
the ECs, the CPS gets insight into the capacity of each
EC as the offered energy is proportional to the available energy. It then optimizes the unit energy price $p_n^*$
for each $n$, within constraint \eqref{eqn-price-constraint-ch-4},
by standard convex optimization. In the second step, each EC $n$
receives the optimized price $p_n^*$ from the CPS, and amends the
offered energy $e_n=e_n^*$ to be supplied to the CPS by playing a
GNEP for the price $p_n^*$. The GNEP, in both steps, reaches the VE
as soon as the slack variables $\xi_n~\forall n\in\mathcal{N}$ reach
the same value $\xi_n=\xi$.
However, the Stackelberg game reaches the EMES when the GNEP amongst the ECs
reaches the VE for the optimized price vector $\mathbf{p^*}$.

In SSHPM, a geometrical interpretation is used and two
projections per iteration are required. Suppose $e^{(k)}$ is the current approximation of the solution of VI$(\mathbf{E};\mathbf{F})$.
First, the projection $r(e^{(k)}) = \text{Proj}_\mathbf{E}[e^{(k)}-\mathbf{F}(e^{(k)})]$ is computed, where $\text{Proj}_\mathbf{E}[z] = \text{argmin}_{w\in \mathbf{E}}\parallel w-z\parallel, w\in \mathbb{R}$. Then, a point $z^{(k)}$ is searched in the line segment between $e^{(k)}$ and $r(e^{(k)})$ such that the hyperplane $\partial H \triangleq \{e\in \mathbb{R}|\langle\mathbf{F}(z^{(k)}), e-z^{(k)}\rangle=0\}$ strictly separates $e^{(k)}$ from any solution $e^*$ of the problem. Once the hyperplane is constructed, $e^{(k+1)}$ is computed in the next iteration onto the intersection of feasible set $\mathbf{E}$ with the hyperspace $H^{(k)} \triangleq \{e\in \mathbb{R}|\langle\mathbf{F}(z^{(k)}), e-z^{(k)}\rangle\leq0\}$ which contains the solution set. Further details on the implementation of SSHPM can be found in \cite{2012IEEE_JTSG_Tushar}.

Convergence of the proposed algorithm is formally stated in the following proposition.
\begin{proposition}
The proposed algorithm using the hyperplane projection method always converges to the optimal solution.
\end{proposition}
\begin{IEEEproof}
The hyperplane projection
method is always guaranteed to converge to a non-empty solution if the problem is strongly monotone~\cite{2007-J4OR_Facchinei}, which
is the case for the proposed algorithm. Furthermore, for the energy amount offered by
the ECs, the optimization problem of the CPS also always converges
to a unique solution due to its strict convexity. Thus, the proposed
algorithm is guaranteed to converge to an optimal solution for the given
constraints in \eqref{eqn-supply-constraint-ch-4} and
\eqref{eqn-price-constraint-ch-4}.
\end{IEEEproof}

\section{Numerical Results}\label{numerical-results-ch-4}
% ------------------------------------------------------------------
% ------------------------------------------------------------------
We consider an example in which a number of ECs are participating in energy trading with the CPS, which has an energy deficiency in a time slot of interest. %We assume that each EC represents $20$ similar energy entities. The smallest energy entity is a smart home equipped with a solar array with $3.6$ kilowatt-hour (kWh) battery capacity, and the largest one is a wind farm with a group of RPI$080$ wind turbines, each with rated capacity around $12.25$ kWh~\cite{2010-ReDriven}.
The available energy of any EC is assumed to be a
uniformly distributed random variable in the range of $\left[64,
240\right]$ kWh. %representing an average unit energy price $37$ US cents for each EC which is a standard electricity tariff in the U.S.A.~\cite{2012-solarchoice}.
Other parameters are chosen as $E_{\text{def}}=700$ kWh, $P=185$ US cents per KWh, $r=2$, $c_n=0.5$, $p_{\text{max}}=P$, $p_{\text{min}}=8.45~\cite{2012-solarchoice}$ and $a_n=1, b_n=1$ for all $n$, unless stated otherwise. Note that the other costs in the total purchasing cost, such as the one associated with insufficient energy purchasing, are not considered in the simulation. Should these costs be accounted for, $P$ needs to be carefully determined in relation to them.
All results are averaged over all possible random values of the ECs' capacities, using $1000$ independent simulation runs, and no anomaly is observed, such as failing to produce a solution, in any iteration.

%\subsection{Convergence to the equilibrium}
% -----------------------------------------------------------------
% ------------------------------------------------------------------

\begin{figure}[t!]
\centering
\subfigure[Convergence of the utility of each EC.]
{\includegraphics[width=\columnwidth]{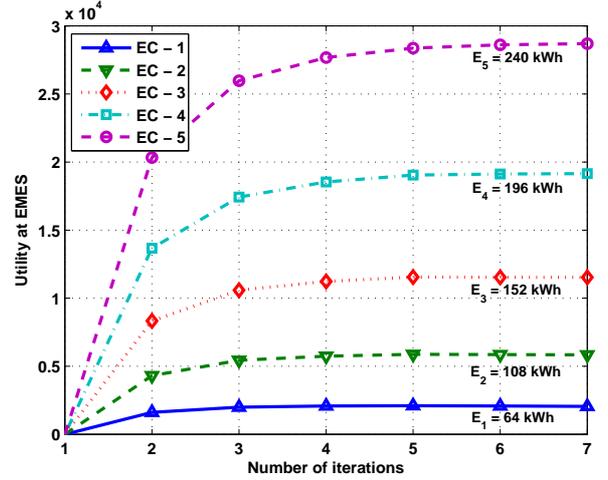}
\label{fig-EMES-utility-ch-4}}
\subfigure[Convergence of the energy supplied by each EC]
{\includegraphics[width=\columnwidth]{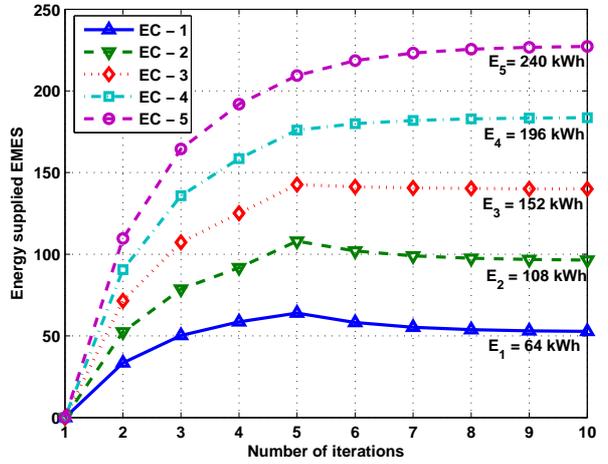}
\label{fig-EMES-energy-ch-4}}
\subfigure[Convergence of the cost to the CPS.]
{\includegraphics[width=\columnwidth]{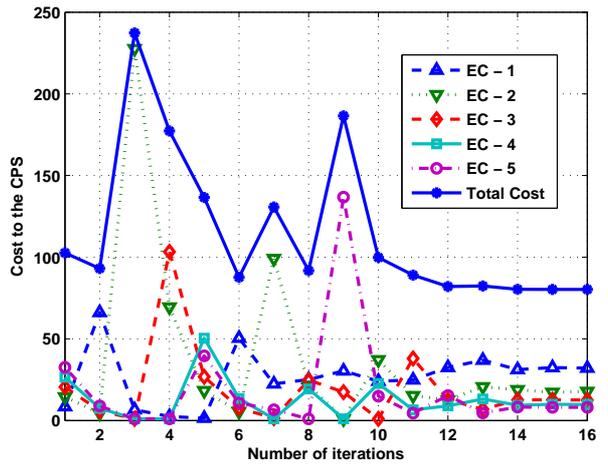}
\label{fig-EMES-cost-ch-4}}
\caption{Convergence to the EMES for the ECs and CPS.}
\label{fig-EMES-convergence}
\vspace{-5pt}
\end{figure}

Fig.~\ref{fig-EMES-convergence} demonstrates the convergence of the
utility achieved by each EC, the amount of energy sold by each EC, and the cost incurred by the CPS during the energy trading process in a random simulation. In this example, the energy deficiency is $E_{\text{def}}=700$ kWh, and $5$ ECs are considered and the randomly generated values of the available energy are depicted as $E_1$ to $E_5$ in the figure. From Fig.~\ref{fig-EMES-utility-ch-4} and
Fig.~\ref{fig-EMES-energy-ch-4}, we can see that both the utility and offered energy for each EC linearly increase with iterations increasing, and utility and offered energy increase towards equilibrium in a similar fashion. An EC with more available energy sells more and achieves higher utility. Both the offered
energy and the achieved utility converge to the EMES after approximately $6$ iterations. Fig.~\ref{fig-EMES-cost-ch-4} shows the variation of the unit energy price determined by CPS during the trading. Unlike the energy and utility curves which almost increase monotonically in iterations, the unit energy price fluctuates a lot, until it reaches the EMES. Fig.~\ref{fig-EMES-cost-ch-4} also clearly show that discriminate unit energy prices are achieved at the EMES, validating one of the goals of the proposed scheme. ECs have less energy for sale are offered higher unit energy price, to be motivated to participate in the energy trading.

\begin{figure}[t!]
\centering
\subfigure[Effects of energy deficiency and number of ECs on utility]
{\includegraphics[width=\columnwidth]{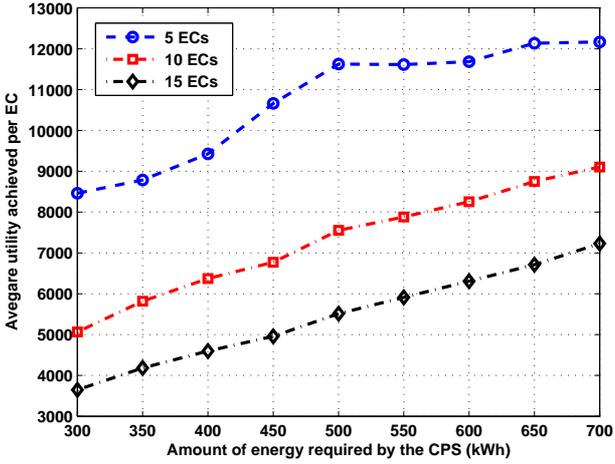}
\label{fig-avg-utility-ch-4}}
\subfigure[Effects of number of ECs on total cost]
{\includegraphics[width=\columnwidth]{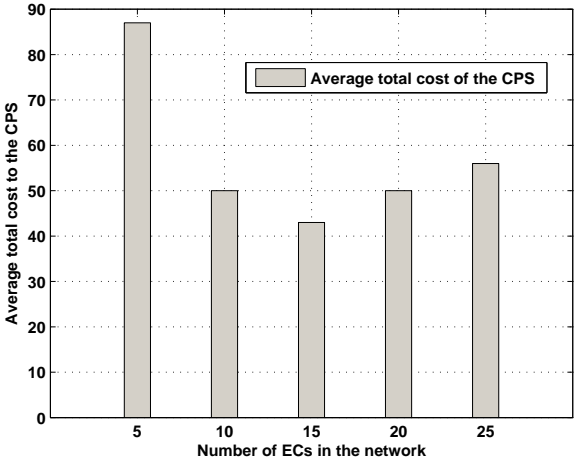}
\label{fig-avg-costVsAES-ch-4}}
\caption{Effects of varying number of ECs on the scheme.}
\label{fig-avg-utility}
\vspace{-5pt}
\end{figure}

%\subsection{Effects on EC's utility and cost to CPS}
In Fig. \ref{fig-avg-utility}, we demonstrate the effects of the number of UEs on the proposed scheme. Fig.~\ref{fig-avg-utility-ch-4} shows how the total energy required by the CPS affects the average utility achieved by each EC, for $5$, $10$ and $15$ ECs. The average utility achieved by each EC decreases with an increasing number of ECs, but increases consistently with increasing energy deficiency. This demonstrates the robustness of the proposed scheme. Fig.~\ref{fig-avg-costVsAES-ch-4} shows how the average total cost to the CPS is affected by the number of UEs, when $E_{\text{def}}=700$ kWh. Interestingly, the total cost incurred by the CPS gradually decreases as the number of ECs increases from $5$ to $15$, and starts increasing with an increase in ECs from $20$ to $25$. In fact, for a fixed price, increasing the number of ECs from $5$ to $15$ allows the CPS to buy its required energy from more ECs at a lower price and consequently the total cost gradually decreases. However, the CPS needs to pay at least the minimum amount (here $p_\text{min}=8.45$ cents/kWh) to each customer to keep it trading energy. Hence, as the number of ECs increases from $20$ to $25$, the total cost increases due to this mandatory minimum payment to more ECs in the network.
%
%\subsection{Effect of the number of ECs on average utility and cost}
%\label{effect-AES-cost-ch-4}
% ---------------------------------------------------------------------------

\begin{figure}[t!]
\centering
{\includegraphics[width=\columnwidth]{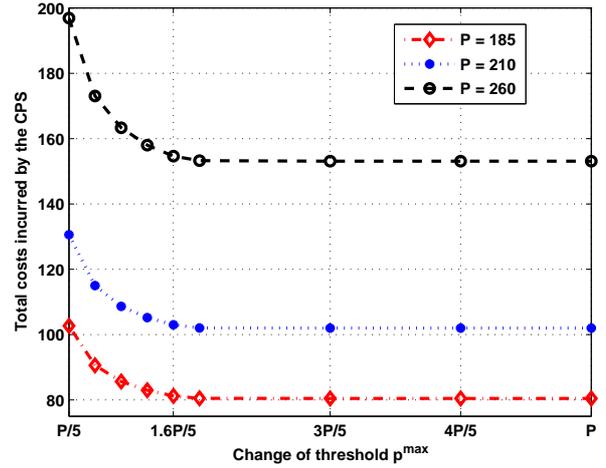}
\caption{Effect of the threshold $p^\text{max}$ on the average total cost to the CPS.}
\label{fig-cost-threshold-ch-4}}
%\label{fig-avg-total-cost}
\vspace{-5pt}
\end{figure}

Fig.~\ref{fig-cost-threshold-ch-4} illustrates how the total cost is affected by the total unit price $P$ and the price upper bound $p_{\text{max}}$, where $N=5$. We assume that the CPS can pay a
maximum of between $P/N$ and $P$ cents per kWh to any EC. As can be seen from the figure, the average total cost
incurred by the CPS eventually decreases as $p_{\text{max}}$ increases, and then reaches a stable state immune to any price change. In fact keeping the threshold at $P/N$ restricts the freedom of the CPS in choosing its unit energy price from any EC, and consequently it incurs a higher total cost. As $p_{\text{max}}$ increases the CPS can choose a higher price, bounded by $p_{\text{max}}$, to pay to the EC with less energy, which in turn enables the CPS to pay a lower price to other ECs in the network, and consequently the total cost to the CPS decreases. Nevertheless, at a particular threshold, the CPS can minimize its own cost by price optimization, and hence there is no change in average total cost with further change in $p_{\text{max}}$. The figure also shows that the total cost is proportional to $P$ and the difference between different cost for different $P$ almost remains as a constant when $p_{\text{max}}$ varies, which is consistent with the analytical results in Section \ref{sec-cpsdec}.%
%Furthermore, as the total price per unit energy increases, the total cost incurred by the CPS increases as expected.

To show the effectiveness of the proposed scheme, we compare it with a standard FIT scheme~\cite{2010-NREL_Couture}. An FIT scheme is a long-term incentive based energy trading scheme designed to encourage the uptake of
renewable energy systems that provide the main grid with power, e.g., when the grid does not have enough supply to meet demand. A higher tariff is paid to the electricity producers as an incentive to take part in the FIT scheme. For
comparison, it is assumed that the contract between the energy
sources and the CPS is such that the sources are capable of
providing the energy the CPS requires. For the FIT scheme, the per
unit tariff is considered to be $60$ US cents/kWh~\cite{2012-solarchoice}.

\begin{figure}[t!]
\centering
\subfigure[Average utility achieved by each EC]
{\includegraphics[width=\columnwidth]{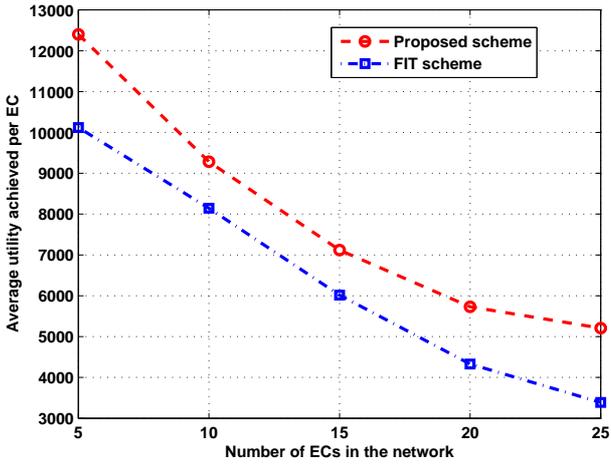}
\label{fig-1-comparison-ch-4}}
\subfigure[Comparison of  cost to the CPS with respect to total per-unit price.]
{\includegraphics[width=\columnwidth]{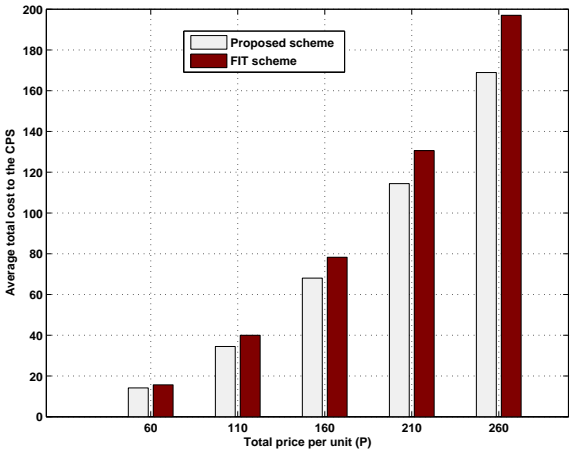}
\label{fig-3-comparison-ch-4}}
\caption{Comparisons of utility, cost and price, between the proposed scheme and the Feed-in-Tariff scheme.}
\vspace{-5pt}
\end{figure}

In \cite{2013-CICC_Tushar}, we studied the performance comparison between the proposed scheme and the FIT scheme based on the average total cost to the CPS for different network sizes. We showed that for a smaller size network of $5$ to $15$ ECs, the proposed scheme has significantly lower cost than the FIT scheme. However, as the network size increases, due to the mandatory payment to a large number of ECs, the cost for the proposed scheme becomes closer to that of the FIT scheme. Here, we compare the average utility per EC for various network sizes, and the average total cost to the CPS as the total unit energy price changes in Fig.~\ref{fig-1-comparison-ch-4} and Fig.~\ref{fig-3-comparison-ch-4}  respectively. Fig.~\ref{fig-1-comparison-ch-4} shows that, as the number
of ECs increases in the network, the average utility
reduces for both schemes. However, the
utility for the proposed scheme is always shown to be better than
the utility achieved by the ECs for the FIT scheme. This is due to
the fact that the proposed scheme allocates the amount of energy for
each EC, using a Stackelberg game, in such a way that the consumer's benefit is
maximized. In contrast, the FIT is a contract based scheme that
makes the customers supply the amount stipulated in their contracts
irrespective of the current
situation. As
shown in Fig.~\ref{fig-1-comparison-ch-4}, for the proposed scheme each EC in the network
achieves an averaged utility $1.5$ times better than that achieved by adopting
the FIT scheme, where the number $1.5$ is obtained by averaging over all different sets, i.e., 5, 10, 15, 20 and 25, of ECs studied in the system.
Assuming the same total price per unit energy for both the proposed
and the FIT schemes, the change in the average total cost to the CPS
for buying energy from the ECs is shown in
Fig.~\ref{fig-3-comparison-ch-4} to increase in proportion to the increase in total price per unit of
energy $P$, as explained for
Fig.~\ref{fig-cost-threshold-ch-4}. However, due to the optimal
allocation of $P$ for each EC, the average total cost for
the proposed scheme is always lower than that of the FIT scheme. The performance benefit of the proposed scheme is
also shown to increase with increasing $P$. This is due to price optimization by the CPS
of the proposed scheme in response to the current VE energy demand of the
ECs, in contrast with the contract-based payment of the FIT scheme.
\vspace{-5pt}
\section{Conclusion}\label{conclusion-ch-4}
In this paper, a consumer-centric energy management scheme for smart
grids has been studied, which is based on maximizing end-user benefits,
as well as keeping the total cost to the central power station at a minimum. Novel utility and cost models are proposed, and a Stackleberg game is formulated to solve the optimization problem. It is shown that the game reaches a Stackelberg equilibrium, which consists of the socially optimal energy and price vector for the ECs and the CPS respectively. The properties of the solution have also been studied. Moreover, a decentralized algorithm has been proposed that can be
implemented by the energy consumers and the central power station with limited communication requirements. The effectiveness of the scheme has been demonstrated via simulation, with noticeable performance improvements over a conventional feed-in-tariff scheme.

The proposed scheme can be extended and improved in various aspects. The constants in the system model can be better calibrated using practical usage data. The interaction of different parameters in the system model is worthy of further investigation, according to the preliminary, but already very interesting, simulation results disclosed in this paper. One limitation of the per time slot based approach is that it ignores the fact that a predominant source of demand side flexibility stems from inter-temporal elasticity of substitution. The proposed scheme can be improved to treat this problem by introducing learning curves for key parameters such as $P$, $E_{\text{def}}$, $r$ and $c_n$. Dependence between inter-temporal parameters can also be described by state equations, which can be formulated according to approaches proposed in \cite{Nie06}. The state equations may define the current state of the system, e.g., $E_{\text{def}}$ and $P$ at the current time slots, as a function of other parameters such as $P$ and $\sum{e_n}$ at the previous time slot. By introducing a learning capability and dependency for key system parameters, the proposed scheme can be extended to efficiently characterize the inter-temporal behavior of the energy management problem.
\def\baselinestretch{.94}
%\bibliographystyle{IEEEtran}
%\bibliography{Thesis_reference}
% --------

% ---------
\begin{IEEEbiography}[{\includegraphics[width=1in,height=1.25in,clip,keepaspectratio]{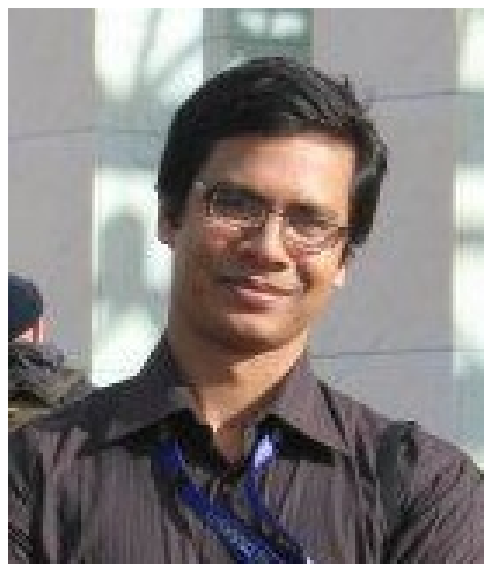}}]{Wayes Tushar}
received the B.Sc. degree in Electrical and Electronic Engineering from Bangladesh University of Engineering and Technology (BUET), Bangladesh, in 2007 and the Ph.D. degree in Engineering from the Australian National University (ANU), Australia in 2013.

Currently, he is a postdoctoral research fellow at Singapore University of Technology and Design (SUTD), Singapore. Prior joining SUTD, he was a visiting researcher at National ICT Australia (NICTA) in ACT, Australia. He was also a visiting student research collaborator in the School of Engineering and Applied Science at Princeton University, NJ, USA during summer 2011. His research interest includes signal processing for distributed networks, game theory and energy management for smart grids. He is the recipient of two best paper awards, both as the first author, in Australian Communications Theory Workshop (AusCTW), 2012 and IEEE International Conference on Communications (ICC), 2013.
\end{IEEEbiography}
\begin{IEEEbiography}[{\includegraphics[width=1in,height=1.25in,clip,keepaspectratio]{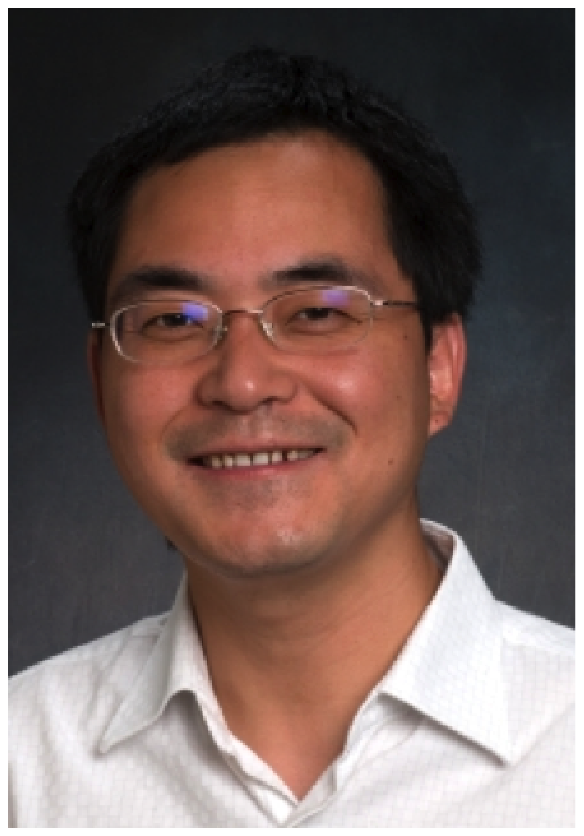}}]{Jian (Andrew) Zhang}(M'04, SM'11) received the B.S. degree from Xi'an JiaoTong University, China, in 1996, the M.Sc. degree from Nanjing University of Posts and Telecommunications, China, in 1999, and the Ph.D. degree from the Australian National University, in 2004. 

Currently, he is a senior research scientist in CSIRO Computational Informatics, Sydney, Australia. From 1999 to 2001, he was a system and hardware engineer in ZTE Corp., Nanjing, China. From 2004 to 2010, he was a researcher in the Networked Systems, NICTA,  Australia. He is an adjunct honorary fellow in the Macquarie University and University of South Australia. His research interests are in the area of signal processing for wireless communications, with focus on MIMO, multicarrier, ultra wideband and sensor networks. He has published 70 papers on leading international Journals and conferences. He is a recipient of CSIRO Chairman's Medal and the Australian Engineering Innovation Award in 2012 for exceptional research achievements in multi-gigabit wireless communications.
\end{IEEEbiography}
\begin{IEEEbiography}[{\includegraphics[height=1.25in]{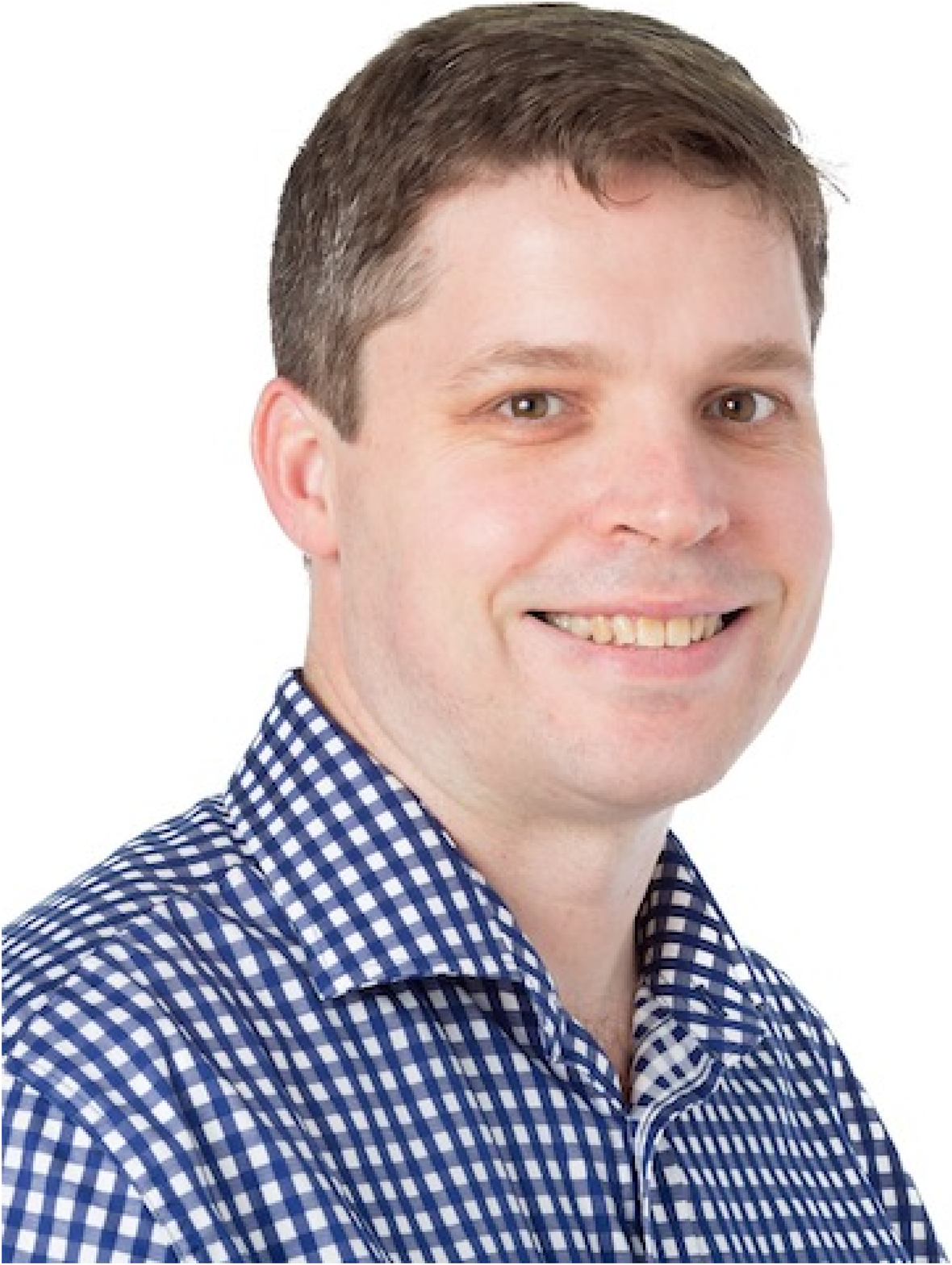}}]{David Smith} is a Senior Researcher at National ICT Australia (NICTA) and is an adjunct Fellow with the Australian National University (ANU), and has been with NICTA and the ANU since 2004. He received the B.E. degree in Electrical Engineering from the University of N.S.W. Australia in 1997, and while studying toward this degree he was on a CO-OP scholarship. He obtained an M.E. (research) degree in 2001 and a Ph.D. in 2004 both from the University of Technology, Sydney (UTS), and both in Telecommunications Engineering. His research interests are in technology and systems for wireless body area networks; game theory for distributed networks; mesh networks; disaster tolerant networks; radio propagation and electromagnetic modeling; MIMO wireless systems; coherent and non-coherent space-time coding; and antenna design, including the design of smart antennas. He also has research interest in distributed optimization for smart grid. He has also had a variety of industry experience in electrical engineering; telecommunications planning; radio frequency, optoelectronic and electronic communications design and integration. He has published over 70 technical refereed papers and made various contributions to IEEE standardization activity; and has received four conference best paper awards.
\end{IEEEbiography}
\begin{IEEEbiography}[{\includegraphics[width=1in,height=1.25in,clip,keepaspectratio]{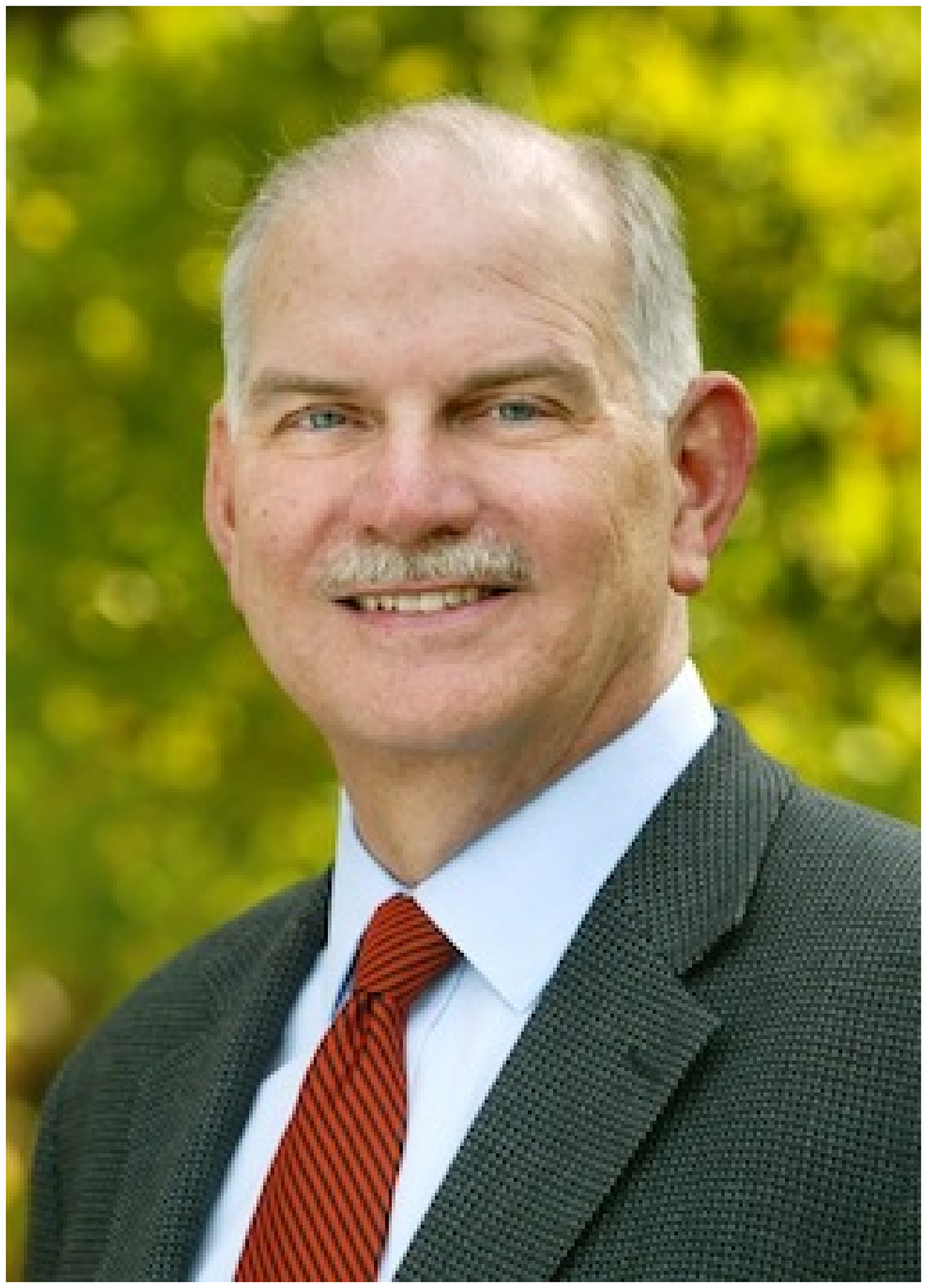}}]{H. Vincent Poor}
(S'72, M'77, SM'82, F'87) received the Ph.D. degree in EECS from Princeton University in 1977.  From 1977 until 1990, he was on the faculty of the University of Illinois at Urbana-Champaign. Since 1990 he has been on the faculty at Princeton, where he is the Michael Henry Strater University Professor of Electrical Engineering and Dean of the School of Engineering and Applied Science. Dr. Poor's research interests are in the areas of stochastic analysis, statistical signal processing, and information theory, and their applications in wireless networks and related fields such as social networks and smart grid. Among his publications in these areas are the recent books \emph{Smart Grid Communications and Networking} (Cambridge University Press, 2012) and \emph{Principles of Cognitive Radio} (Cambridge University Press, 2013).

Dr. Poor is a member of the U. S. National Academy of Engineering, the U. S. National Academy of Sciences, and Academia Europaea. He is also a Fellow of the American Academy of Arts and Sciences, an International Fellow of the Royal Academy of Engineering (U.K.), and a Corresponding Fellow of the Royal Society of Edinburgh.   In 1990, he served as President of the IEEE Information Theory Society, and in 2004-07 he served as the Editor-in-Chief of the \emph{IEEE Transactions on Information Theory}. He received a Guggenheim Fellowship in 2002 and the IEEE Education Medal in 2005. Recent recognition of his work includes the 2010 IET Ambrose Fleming Medal, the 2011 IEEE Eric E. Sumner Award, the 2011 Society Award of the IEEE Signal Processing Society, and honorary doctorates from Aalborg University, the Hong Kong University of Science and Technology and the University of Edinburgh.  
\end{IEEEbiography}
\begin{IEEEbiography}[{\includegraphics[height=1.25in]{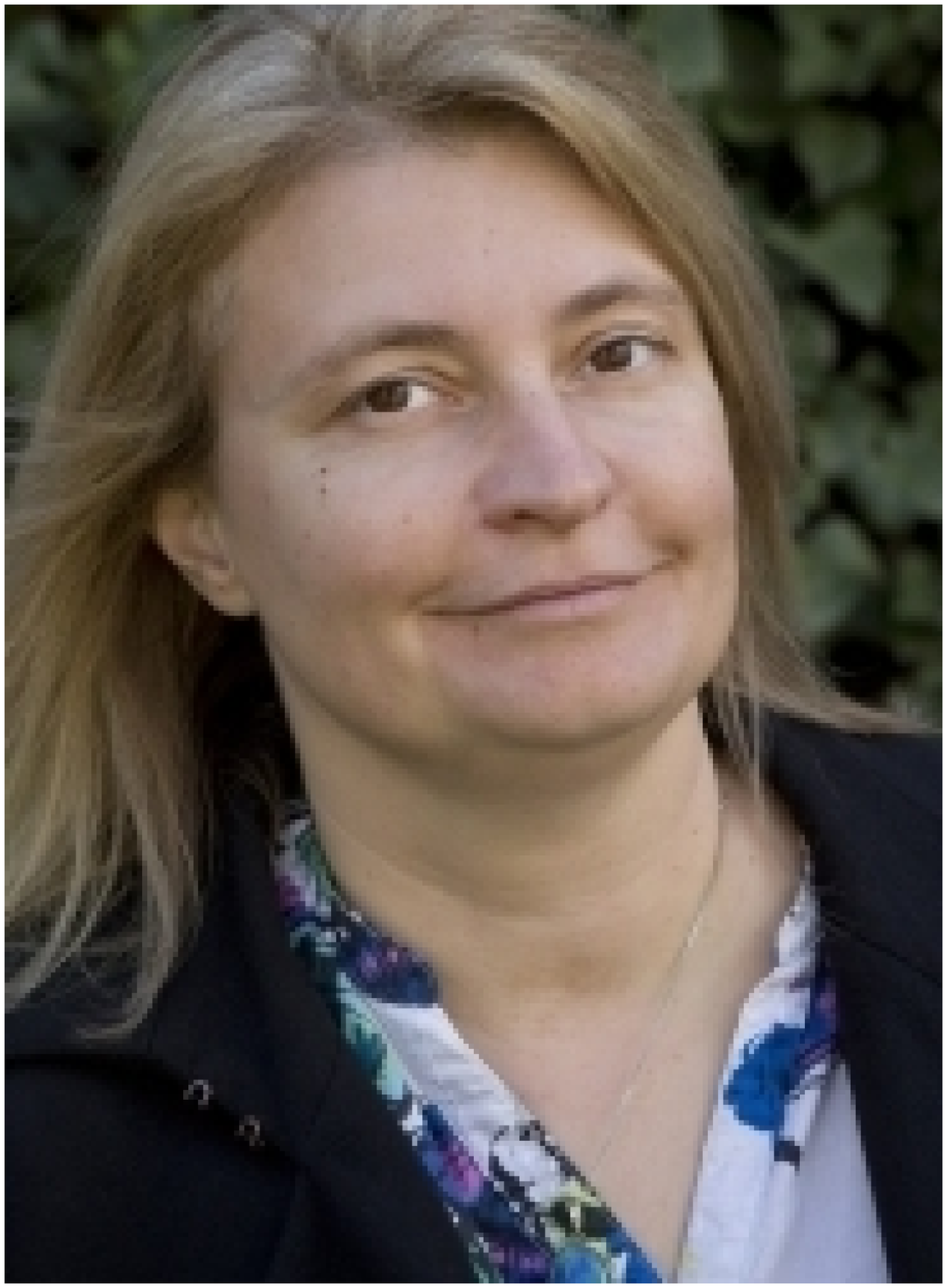}}]{Sylvie Thi{\'{e}}baux} is a professor of computer science at the Australian
National University and a research leader in the Optimisation Research
Group at NICTA, where she heads the Future Energy Systems project. She
received a PhD in computer science from the university of Rennes in
1995. Before joining the ANU, she held research appointments with the
national research centers INRIA in France and with CSIRO in
Australia. In the recent past she was the director of NICTA's Canberra
Laboratories, home to 150 researchers and PhD students.  Her research
interests are in artificial intelligence and optimisation, in
particular automated planning and scheduling, model-based diagnosis,
combinatorial optimisation and search, reasoning under uncertainty,
and their applications to energy and transport.  She is an associate
editor of the Artificial Intelligence journal (AIJ), the president of the 
International Conference on Automated Planning and Scheduling (ICAPS),
and a Councillor of the Association for the Advancement of Artificial
Intelligence (AAAI). 
\end{IEEEbiography}
\end{document}